\pdfoutput=1
\documentclass[12pt,a4paper]{article}

\usepackage{ifthen} 
\newboolean{pdflatex}
\setboolean{pdflatex}{true} 

\newboolean{articletitles}
\setboolean{articletitles}{true} 

\newboolean{uprightparticles}
\setboolean{uprightparticles}{false} 

\newboolean{inbibliography}
\setboolean{inbibliography}{false} 

\usepackage{microtype}
\usepackage{lineno}  
\usepackage{xspace} 
\usepackage{caption} 

\usepackage{graphicx}  
\usepackage{color}
\usepackage{colortbl}
\graphicspath{{./figs/}} 

\usepackage{amsmath} 
\usepackage{amssymb}
\usepackage{amsfonts}
\usepackage{upgreek} 

\newcommand*\patchAmsMathEnvironmentForLineno[1]{%
\expandafter\let\csname old#1\expandafter\endcsname\csname #1\endcsname
\expandafter\let\csname oldend#1\expandafter\endcsname\csname
end#1\endcsname
 \renewenvironment{#1}%
   {\linenomath\csname old#1\endcsname}%
   {\csname oldend#1\endcsname\endlinenomath}%
}
\newcommand*\patchBothAmsMathEnvironmentsForLineno[1]{%
  \patchAmsMathEnvironmentForLineno{#1}%
  \patchAmsMathEnvironmentForLineno{#1*}%
}
\AtBeginDocument{%
\patchBothAmsMathEnvironmentsForLineno{equation}%
\patchBothAmsMathEnvironmentsForLineno{align}%
\patchBothAmsMathEnvironmentsForLineno{flalign}%
\patchBothAmsMathEnvironmentsForLineno{alignat}%
\patchBothAmsMathEnvironmentsForLineno{gather}%
\patchBothAmsMathEnvironmentsForLineno{multline}%
}

\usepackage{hyperref}    
\usepackage[all]{hypcap} 

\def\lhcb {\mbox{LHCb}\xspace}

\def\babar  {\mbox{BaBar}\xspace}
\def\belle  {\mbox{Belle}\xspace}

\def\dzero  {\mbox{D0}\xspace}

\ifthenelse{\boolean{uprightparticles}}%
{

 \def\Pmu         {\ensuremath{\upmu}\xspace}                 
 \def\Pnu         {\ensuremath{\upnu}\xspace}                 
                  
 \def\Ppi         {\ensuremath{\uppi}\xspace}

 \def\Pphi        {\ensuremath{\upphi}\xspace}

 \def\Ppsi        {\ensuremath{\uppsi}\xspace}

 \def\PDelta      {\ensuremath{\Delta}\xspace}                 
 \def\PXi      {\ensuremath{\Xi}\xspace}                 
 \def\PLambda      {\ensuremath{\Lambda}\xspace}                 
 \def\PSigma      {\ensuremath{\Sigma}\xspace}                 
 \def\POmega      {\ensuremath{\Omega}\xspace}                 
 \def\PUpsilon      {\ensuremath{\Upsilon}\xspace}                 
 
 \def\PB      {\ensuremath{\mathrm{B}}\xspace}                 
                  
 \def\PD      {\ensuremath{\mathrm{D}}\xspace}

 \def\PJ      {\ensuremath{\mathrm{J}}\xspace}                 
 \def\PK      {\ensuremath{\mathrm{K}}\xspace}

 \def\PP      {\ensuremath{\mathrm{P}}\xspace}

 \def\Pd      {\ensuremath{\mathrm{d}}\xspace}

 \def\Pi      {\ensuremath{\mathrm{i}}\xspace}

 \def\Ps      {\ensuremath{\mathrm{s}}\xspace}

 \def\Pz      {\ensuremath{\mathrm{z}}\xspace}                 
}
{

 \def\Pmu         {\ensuremath{\mu}\xspace}                 
 \def\Pnu         {\ensuremath{\nu}\xspace}                 
                  
 \def\Ppi         {\ensuremath{\pi}\xspace}

 \def\Pphi        {\ensuremath{\phi}\xspace}

 \def\Ppsi        {\ensuremath{\psi}\xspace}                 
                  
 \mathchardef\PDelta="7101
 \mathchardef\PXi="7104
 \mathchardef\PLambda="7103
 \mathchardef\PSigma="7106
 \mathchardef\POmega="710A
 \mathchardef\PUpsilon="7107
                  
 \def\PB      {\ensuremath{B}\xspace}                 
                  
 \def\PD      {\ensuremath{D}\xspace}

 \def\PJ      {\ensuremath{J}\xspace}                 
 \def\PK      {\ensuremath{K}\xspace}

 \def\PP      {\ensuremath{P}\xspace}

 \def\Pd      {\ensuremath{d}\xspace}

 \def\Pi      {\ensuremath{i}\xspace}

 \def\Ps      {\ensuremath{s}\xspace}

 \def\Pz      {\ensuremath{z}\xspace}                 
}

\makeatletter
\ifcase \@ptsize \relax
  \newcommand{\miniscule}{\@setfontsize\miniscule{4}{5}}
\or
  \newcommand{\miniscule}{\@setfontsize\miniscule{5}{6}}
\or
  \newcommand{\miniscule}{\@setfontsize\miniscule{5}{6}}
\fi
\makeatother

\DeclareRobustCommand{\optbar}[1]{\shortstack{{\miniscule (\rule[.5ex]{1.25em}{.18mm})}
  \\ [-.7ex] $#1$}}

\def\mup        {{\ensuremath{\Pmu^+}}\xspace}

\def\neu        {{\ensuremath{\Pnu}}\xspace}

\def\neum       {{\ensuremath{\neu_\mu}}\xspace}

\def\dquark    {{\ensuremath{\Pd}}\xspace}

\def\squark    {{\ensuremath{\Ps}}\xspace}

\def\pion   {{\ensuremath{\Ppi}}\xspace}
\def\piz    {{\ensuremath{\pion^0}}\xspace}

\def\pip    {{\ensuremath{\pion^+}}\xspace}
\def\pim    {{\ensuremath{\pion^-}}\xspace}

\def\kaon    {{\ensuremath{\PK}}\xspace}
  \def\Kbar    {{\kern 0.2em\overline{\kern -0.2em \PK}{}}\xspace}

\def\KorKbar    {\kern 0.18em\optbar{\kern -0.18em K}{}\xspace}
\def\Kz      {{\ensuremath{\kaon^0}}\xspace}

\def\Kp      {{\ensuremath{\kaon^+}}\xspace}
\def\Km      {{\ensuremath{\kaon^-}}\xspace}

\def\KS      {{\ensuremath{\kaon^0_{\rm\scriptscriptstyle S}}}\xspace}
\def\KL      {{\ensuremath{\kaon^0_{\rm\scriptscriptstyle L}}}\xspace}

  \def\Dbar    {{\kern 0.2em\overline{\kern -0.2em \PD}{}}\xspace}
\def\D       {{\ensuremath{\PD}}\xspace}

\def \DorDbar   {\kern 0.18em\optbar{\kern -0.18em D}{}\xspace}
\def\Dz      {{\ensuremath{\D^0}}\xspace}

\def\Dsm     {{\ensuremath{\D^-_\squark}}\xspace}

\def\B       {{\ensuremath{\PB}}\xspace}
\def\Bbar    {{\ensuremath{\kern 0.18em\overline{\kern -0.18em \PB}{}}}\xspace}

\def\BorBbar    {\kern 0.18em\optbar{\kern -0.18em B}{}\xspace}
\def\Bz      {{\ensuremath{\B^0}}\xspace}
\def\Bzb     {{\ensuremath{\Bbar{}^0}}\xspace}
\def\Bu      {{\ensuremath{\B^+}}\xspace}

\def\Bp      {{\ensuremath{\Bu}}\xspace}

\def\Bd      {{\ensuremath{\B^0}}\xspace}
\def\Bs      {{\ensuremath{\B^0_\squark}}\xspace}

\def\Bdb     {{\ensuremath{\Bbar{}^0}}\xspace}

\def\jpsi     {{\ensuremath{{\PJ\mskip -3mu/\mskip -2mu\Ppsi\mskip 2mu}}}\xspace}

  \def\Y#1S{\ensuremath{\PUpsilon{(#1S)}}\xspace}

\def\FourS {{\Y4S}}

\def\Lbar        {{\ensuremath{\kern 0.1em\overline{\kern -0.1em\PLambda}}}\xspace}
\def\LorLbar    {\kern 0.18em\optbar{\kern -0.18em \PLambda}{}\xspace}

\def\to                 {\ensuremath{\rightarrow}\xspace}

\def\order   {{\ensuremath{\mathcal{O}}}\xspace}

\def\CP                {{\ensuremath{C\!P}}\xspace}
\def\CPT               {{\ensuremath{C\!PT}}\xspace}

\newcommand{\dms}{{\ensuremath{\Delta m_{\squark}}}\xspace}
\newcommand{\dmd}{{\ensuremath{\Delta m_{\dquark}}}\xspace}
\newcommand{\DG}{{\ensuremath{\Delta\Gamma}}\xspace}
\newcommand{\DGs}{{\ensuremath{\Delta\Gamma_{\squark}}}\xspace}
\newcommand{\DGd}{{\ensuremath{\Delta\Gamma_{\dquark}}}\xspace}

\newcommand{\Delm}{{\mbox{$\Delta m $}}\xspace}

\def\AT#1     {\ensuremath{A_{\mathrm{T}}^{#1}}\xspace}           

\def\C#1      {\ensuremath{\mathcal{C}_{#1}}\xspace}                       
\def\Cp#1     {\ensuremath{\mathcal{C}_{#1}^{'}}\xspace}                    
\def\Ceff#1   {\ensuremath{\mathcal{C}_{#1}^{\mathrm{(eff)}}}\xspace}        
\def\Cpeff#1  {\ensuremath{\mathcal{C}_{#1}^{'\mathrm{(eff)}}}\xspace}       
\def\Ope#1    {\ensuremath{\mathcal{O}_{#1}}\xspace}                       
\def\Opep#1   {\ensuremath{\mathcal{O}_{#1}^{'}}\xspace}                    

\newcommand{\ket}[1]{\ensuremath{|#1\rangle}}              

\newcommand{\tev}{\ifthenelse{\boolean{inbibliography}}{\ensuremath{~T\kern -0.05em eV}\xspace}{\ensuremath{\mathrm{\,Te\kern -0.1em V}}}\xspace}
\newcommand{\gev}{\ensuremath{\mathrm{\,Ge\kern -0.1em V}}\xspace}
\newcommand{\mev}{\ensuremath{\mathrm{\,Me\kern -0.1em V}}\xspace}
\newcommand{\kev}{\ensuremath{\mathrm{\,ke\kern -0.1em V}}\xspace}
\newcommand{\ev}{\ensuremath{\mathrm{\,e\kern -0.1em V}}\xspace}
\newcommand{\gevc}{\ensuremath{{\mathrm{\,Ge\kern -0.1em V\!/}c}}\xspace}
\newcommand{\mevc}{\ensuremath{{\mathrm{\,Me\kern -0.1em V\!/}c}}\xspace}
\newcommand{\gevcc}{\ensuremath{{\mathrm{\,Ge\kern -0.1em V\!/}c^2}}\xspace}
\newcommand{\gevgevcccc}{\ensuremath{{\mathrm{\,Ge\kern -0.1em V^2\!/}c^4}}\xspace}
\newcommand{\mevcc}{\ensuremath{{\mathrm{\,Me\kern -0.1em V\!/}c^2}}\xspace}

\def\invfb   {\ensuremath{\mbox{\,fb}^{-1}}\xspace}

\def\ps   {\ensuremath{{\rm \,ps}}\xspace}

\def\invps{\ensuremath{{\rm \,ps^{-1}}}\xspace}

\def\order{{\ensuremath{\cal O}}\xspace}

\def\gsim{{~\raise.15em\hbox{$>$}\kern-.85em
          \lower.35em\hbox{$\sim$}~}\xspace}
\def\lsim{{~\raise.15em\hbox{$<$}\kern-.85em
          \lower.35em\hbox{$\sim$}~}\xspace}

\newcommand{\mean}[1]{\ensuremath{\left\langle #1 \right\rangle}} 

\def\degrees{\ensuremath{^{\circ}}\xspace}

\def\tell1  {TELL1\xspace}
\def\ukl1   {UKL1\xspace}

\newcommand{\ACP}{\ensuremath{A_{\CP}}\xspace}
\newcommand{\ACPT}{\ensuremath{A_{\CPT}}\xspace}
\newcommand{\ACPTCP}{\ensuremath{A_{\CPT,\CP}}\xspace}

\newcommand{\Adir}{{\ensuremath{A^{\rm dir}}}\xspace}
\newcommand{\Amix}{{\ensuremath{A^{\rm mix}}}\xspace}

\def\T               {{\ensuremath{T}}\xspace}

\def\Re               {{\ensuremath{\rm Re}}\xspace}
\def\Im               {{\ensuremath{\rm Im}}\xspace}

\newcommand{\GeV}{Ge\kern -0.1em V}

\newcommand{\Ap}{\ensuremath{A_P}\xspace}
\newcommand{\Ad}{\ensuremath{A_D}\xspace}

\def\KSL      {{\ensuremath{\kaon^0_{\rm\scriptscriptstyle S,L}}}\xspace}

\def\Dan          {{\ensuremath{\Delta a_0}}\xspace}
\def\Dax          {{\ensuremath{\Delta a_X}}\xspace}
\def\Day          {{\ensuremath{\Delta a_Y}}\xspace}
\def\Daz          {{\ensuremath{\Delta a_Z}}\xspace}
\def\Dat          {{\ensuremath{\Delta a_T}}\xspace}

\def\Daxy         {{\ensuremath{\Delta a_{X,Y}}}\xspace}
\def\Damu         {{\ensuremath{\Delta a_{\mu}}}\xspace}
\def\DamuK        {{\ensuremath{\Delta a^K_{\mu}}}\xspace}
\def\DamuBd       {{\ensuremath{\Delta a^{\Bd}_{\mu}}}\xspace}
\def\DamuBs       {{\ensuremath{\Delta a^{\Bs}_{\mu}}}\xspace}

\newcommand{\gevcNS}{\ensuremath{{\mathrm{Ge\kern -0.1em V\!/}c}}\xspace}

\def\P       {{\ensuremath{\PP}}\xspace}
\def\fbar    {{\ensuremath{\kern 0.18em\overline{\kern -0.18em f}{}}}\xspace}
\def\Abar    {{\ensuremath{\kern 0.18em\overline{\kern -0.18em A}{}}}\xspace}
\def\Pbar    {{\ensuremath{\kern 0.18em\overline{\kern -0.18em \PP}{}}}\xspace}

\def\Pz      {{\ensuremath{\P^0}}\xspace}
\def\Pzb     {{\ensuremath{\Pbar{}^0}}\xspace}

\def\lambdabar{{\ensuremath{\kern 0.18em\overline{\kern -0.18em \lambda}{}}}\xspace}

\usepackage{cite} 
\usepackage{mciteplus}

\begin{document}

\renewcommand{\thefootnote}{\fnsymbol{footnote}}
\setcounter{footnote}{1}

\begin{titlepage}
\pagenumbering{roman}

\begin{raggedleft}
Nikhef-2014-017 \\ 
\today \\ 
\end{raggedleft}

\vspace*{0.4cm}

{\bf\boldmath\huge
\begin{center}
Status and prospects for $C\!PT$ and Lorentz invariance violation searches in neutral meson mixing
\end{center}
}

\vspace*{0.5cm}

\begin{center}
Jeroen van Tilburg and Maarten van Veghel\\
\vspace*{2mm}
{\small \it Nikhef National Institute for Subatomic Physics, Amsterdam, 
  The Netherlands}\\
\end{center}

\vspace{\fill}

\begin{abstract}
  \noindent
An overview of current experimental bounds on $C\!PT$ violation in neutral meson
mixing is given. New values for the $C\!PT$ asymmetry in the $B^0$ and $B_s^0$
systems are deduced from published BaBar, Belle and LHCb results. With dedicated
analyses, LHCb will be able to further improve the bounds on $C\!PT$ violation
in the $D^0$, $B^0$ and $B_s^0$ systems. Since $C\!PT$ violation implies
violation of Lorentz invariance in an interacting local quantum field theory,
the observed $C\!PT$ asymmetry will exhibit sidereal- and boost-dependent
variations.  Such $C\!PT$-violating and Lorentz-violating effects are
accommodated in the framework of the Standard-Model Extension (SME). The large
boost of the neutral mesons produced at LHCb results in a high sensitivity to
the corresponding SME coefficients. For the $B^0$ and $B_s^0$ systems, using
existing LHCb results, we determine with high precision the SME coefficients
that are not varying with sidereal time. With a full sidereal analysis, LHCb
will be able to improve the existing SME bounds in the $D^0$, $B^0$ and $B_s^0$
systems by up to two orders of magnitude.
\end{abstract}
\vspace*{1.0cm}

\vspace{\fill}

\end{titlepage}


\newpage
\setcounter{page}{2}


\renewcommand{\thefootnote}{\arabic{footnote}}
\setcounter{footnote}{0}


\pagestyle{plain} 
\setcounter{page}{1}
\pagenumbering{arabic}


\section{Introduction}
\label{sec:Introduction}

In the weak interaction of the Standard Model, the symmetries under
transformations of charge conjugation ($C$), parity ($P$), and time reversal
($T$) are broken. Nevertheless, the combined \CPT transformation is observed to
be an exact fundamental symmetry of nature. From a theoretical perspective, \CPT
symmetry is required by any Lorentz-invariant, local quantum field theory. Many
experimental searches for \CPT violation have been performed over the last
decades. Interferometry in the particle-antiparticle mixing of neutral mesons is
a natural and very sensitive method to search to deviations from \CPT
invariance. Since most \CPT tests have been performed with neutral kaons,
progress can still be made in the \Dz, \Bd and \Bs systems.

As \CPT violation implies Lorentz violation in an interacting local quantum
field theory~\cite{Greenberg:2002uu}, any \CPT-violating observable must also
break Lorentz invariance. In the framework of the Standard Model
Extension~\cite{Colladay:1996iz,Colladay:1998fq} (SME), spontaneous \CPT
violation and Lorentz invariance violation appear in a low-energy effective
field theory. In this sense, small \CPT-violating effects at low energies
provide a window to the quantum gravity scale~\cite{Liberati:2013xla}. Such
effects are expected to be suppressed by $m^2/M_{\rm Pl}$, with $M_{\rm
  Pl}\approx10^{19}\gev$ the Planck mass and $m$ the relevant low-energy mass,
which depends on the underlying unified theory and possibly ranges from the mass
of the neutral meson to the electroweak
mass~\cite{Kostelecky:1991ak,*Kostelecky:1994rn}. The Lorentz violation
introduces a boost- and direction-dependent variation in the \CPT-violating
parameters. From an experimental point of view, the direction dependence results
in a modulation with the sidereal phase. Such modulations would provide an
unambiguous signature of \CPT violation.

We will show that a high sensitivity to these effects can be obtained by
exploiting the large sample of heavy flavour decays obtained at the \lhcb
experiment, in particular taking advantage of the forward boost of the neutral
mesons. Using published \lhcb results, corresponding to a luminosity of
$1\invfb$, we can already deduce improved constraints on the SME parameters.
Based on naive extrapolations, further improvements are possible with dedicated
analyses on the existing $3\invfb$ data set.

\section{Formalism}

The time evolution of a neutral meson system, \Pz-\Pzb, is governed by an
effective $2\times2$ Hamiltonian ${\bf H}={\bf M}-i{\bf \Gamma}/2$. Following
the notation in Ref.~\cite{Aubert:2004xga}, we write the light and heavy mass
eigenstates, with eigenvalues $m_{L,H}-i\Gamma_{L,H}/2$, as
\begin{align}
  \ket{P_L} &= p \sqrt{1-z} \ket{\Pz} +  q \sqrt{1+z} \ket{\Pzb} \nonumber \\
  \ket{P_H} &= p \sqrt{1+z} \ket{\Pz} -  q \sqrt{1-z} \ket{\Pzb} \ ,
\end{align}
where $p$ and $q$ spawn the eigenvectors under \CPT symmetry and $z$ is the
complex, \CPT-violating parameter. The definition of $z$ is independent of phase
convention~\cite{Kostelecky:2001ff}. The mixing parameters are defined as
$\Delm \equiv m_H-m_L$ and $\DG\equiv \Gamma_H - \Gamma_L$, and the average mass
and decay rate as $m \equiv (M_{11} + M_{22})/2$ and $\Gamma \equiv (\Gamma_{11}
+ \Gamma_{22})/2$. This definition implies that $\DG<0$ for the \Kz, \Bd and \Bs
systems, and $\DG>0$ for the \Dz system in the Standard Model. The
\CPT-violating parameter in \Pz-\Pzb mixing can be written as
\begin{align}
   z = \frac{\delta m - i\delta\Gamma/2}{\Delm - i \DG/2} \ , 
\label{eq:z}
\end{align}
where $\delta m\equiv(M_{11}-M_{22})$ and
$\delta\Gamma\equiv(\Gamma_{11}-\Gamma_{22})$ are the differences of the
diagonal mass and decay rate matrix elements of ${\bf H}$. This equation makes
clear that $z$ is sensitive to small values of $\delta m$ or $\delta\Gamma$ due
to the smallness of $\Delm$ and $\DG$ in neutral meson systems. By measuring the
time-dependent decay rates of an initial \Pz or \Pzb state to a final state $f$
or \fbar, information on $z$ can be obtained. For simplicity, we only consider
\CPT violation in \Pz-\Pzb mixing. Direct \CPT violation is experimentally
difficult to separate from direct \CP-violating effects. In both cases, it
causes a difference in the instantaneous decay amplitudes, i.e., $A_f\neq
\Abar_{\fbar}$, where $A_{f,\fbar}$ ($\Abar_{f,\fbar}$) is the direct decay
amplitude of a \Pz (\Pzb) meson to a final state $f$ or $\fbar$. In the
following, any direct \CP-violating term implicitly includes possible direct
\CPT violation. For a complete expression of the decay rates we refer to
Ref.~\cite{Aubert:2004xga}. Although those equations apply to the more general
case of coherent production of \Bd-\Bdb pairs, we will ignore this additional
complication here and assume incoherent production by setting the decay
amplitude of the tagging particle to either zero or one.

It is instructive to construct an observable \CPT asymmetry
\begin{align}
   \ACPT(t) &\equiv \frac{\Pbar_{\fbar}(t)-P_f(t)}{\Pbar_{\fbar}(t)+P_f(t)} \ ,
\label{eq:ACPT}
\end{align}
where $P_f$ ($\Pbar_{\fbar}$) is the time-dependent decay probability of an
initial \Pz (\Pzb) meson to a final state $f$ (\fbar). For decays to pure
flavour-specific final states (i.e., $A_{\fbar}=\Abar_f=0$), this asymmetry can
be written as
\begin{align}
   \ACPT(t) = \Adir + \frac{2\Re(z)\sinh\Delta\Gamma t/2 - 2\Im(z)\sin\Delm t}
        {(1+|z|^2)\cosh\Delta\Gamma t/2 + (1-|z|^2)\cos\Delm t} \ ,
\label{eq:ACPT_fs}
\end{align}
where the direct \CP asymmetry $\Adir\equiv (|\Abar_{\fbar}|^2-|A_f|^2)/
(|\Abar_{\fbar}|^2+|A_f|^2)$ is assumed to be small. On the other hand,
the \CP asymmetry, defined as
\begin{align}
\ACP(t) \equiv \frac{\Pbar_f(t) - P_{\fbar}(t)} {\Pbar_f(t) + P_{\fbar}(t)} \ ,
\end{align}
and the \CPT asymmetry become equivalent for decays to \CP eigenstates
$f=\fbar$, and their effects become automatically connected. The \CPT or \CP
asymmetry can be written as
\begin{align}
  \ACPTCP(t) =& \big[\Amix/2 + D_f\Re(z)\big]\cosh\DG t/2 -
                \big[ C_f + D_f \Re(z) \big] \cos\Delm t \, + \nonumber \\ 
              & \big[\D_f\Amix/2 + \Re(z)\big] \sinh\DG t/2 +
                \big[S_f - \Im(z) \big] \sin\Delm t \ ,
\label{eq:ACPT_CP}
\end{align}
where $D_f=2\Re(\lambda_f)/(1+|\lambda_f|^2)$, $C_f=(1-|\lambda_f|^2)/
(1+|\lambda_f|^2)$ and $S_f=2\Im(\lambda_f)/(1+|\lambda_f|^2)$. The parameter
$\lambda_f=(q/p) (\Abar_f/A_f)$ is introduced for convenience, and
$\Amix=(1-|q/p|^4)/(1+|q/p|^4)$ describes \CP violation in mixing
only. In the absence of \CP violation in mixing (i.e., $|q/p|=1$), $C_f$ is
equivalent to \Adir. Only leading-order terms in $\lambda_f$ and $z$ are
retained in Eq.~\ref{eq:ACPT_CP}. Comparing Eqs.~\ref{eq:ACPT_fs} and
\ref{eq:ACPT_CP}, it becomes apparent that flavour-specific final states and \CP
eigenstates have different, complementary sensitivities to $\Re(z)$ and
$\Im(z)$. We will come back to this point later.

Up to now we have assumed that $z$ is a constant of nature for each of the four
neutral meson systems. We will refer to this assumption as the classical
approach. In the SME Lagrangian, \CPT-violating and Lorentz-violating terms are
introduced for the fermions with coupling coefficients
$a_\mu$~\cite{Kostelecky:1997mh}. The observable effect is determined by the
contributions from the two valence quarks, $q_1$ and ${\overline q}_2$, in a
meson as $\Damu\simeq a_\mu^{q_1} - a_\mu^{q_2}$, hereby ignoring small effects
from binding and normalization. In the SME approach, the equations above remain
valid, but now $z$ depends on the four-velocity
$\beta^\mu=\gamma(1,\vec{\beta})$ of the neutral meson as
\begin{align}
   z \simeq \frac{\beta^\mu \Damu}{\Delm - i \DG/2} \ .
\label{eq:SMEz}
\end{align}
An overview of experimental bounds on \Damu and other SME parameters is given in
Ref.~\cite{Kostelecky:2008ts}. In the SME, \Damu is required to be
real~\cite{Kostelecky:1999bm}, which implies $\delta\Gamma=0$. The real and
imaginary parts of $z$ then become connected through
\begin{align}
 \Re(z)\DG = 2 \Im(z)\Delm \ .
\label{eq:SMEconstraint}
\end{align}
As we will see later, this constraint has implications for \CPT violation
searches within the SME framework.

In such a search, the four-velocity of the neutral mesons at any time needs to
be determined with respect to fixed stars. A useful reference frame is the
Sun-centred frame defined in Ref.~\cite{Kostelecky:1999bm}. In this
frame, the $Z$-axis is directed North, following the rotation axis of Earth, the
$X$-axis points away from the Sun at the vernal equinox and the $Y$-axis
complements the right-handed coordinate system. For an experiment where the
neutral mesons are produced in a horizontal direction, fixed with respect to the
Earth's coordinate system, the dependence on the four-velocity can be written as
\begin{align}
   \beta^\mu\Damu = \gamma [ \Dan + \beta\Daz\cos\chi + 
     \beta\sin\chi(\Day\sin\Omega\hat{t} + \Dax\cos\Omega\hat{t}\,) ] \ ,
\label{eq:SME_GPS}
\end{align}
where $\Omega$ is the sidereal frequency and $\cos\chi=\cos\theta\cos\lambda$
with $\theta$ the azimuth of the neutral mesons and $\lambda$ the latitude. The
time coordinate $\hat{t}$ is chosen such that the boost direction aligns with
the $X$-axis at $\hat{t}=0$ in the $XY$ projection. We have used the same
convention as in Ref.~\cite{Kostelecky:1999bm}, where the spatial coordinates of
the \Damu field are defined such that $\Delta a^{X,Y,Z}=-\Delta
a_{X,Y,Z}$. Equation~\ref{eq:SME_GPS} makes clear that $z$ not only depends on
the size of the boost, but also that it has a constant component, independent of
the sidereal phase, and a component that exhibits a sidereal modulation. The
sidereal variation is largest when the experiment is oriented east-west or when
it is close to the North Pole. For the \lhcb experiment, we determine the
latitude $\lambda=46.24\degrees$N and azimuth $\theta=236.3\degrees$ east of
north, which gives $\cos\chi=-0.38$ and $\sin\chi=0.92$. This means that the
constant component scales with $(\Dan-0.38\Daz)$ and that the sidereal variation
at \lhcb is close to maximal.

\section{Experimental results and potential measurements}

In the following, we present an overview of experimental searches for \CPT
violation in the four neutral meson systems. We interpret published results that
are sensitive to \CPT violation. These new values are summarized in
Table~\ref{tab:CPT_overview} and discussed in the following. We also include
prospects for analyses that can be conducted with current data from the \lhcb
experiment. The expected sensitivities on the \CPT-violating parameters with the
existing $3\invfb$ data set are given in Table~\ref{tab:CPT_reach}.

\renewcommand{\arraystretch}{1.1}
\begin{table}[!b]
  \begin{center}
    \caption{Overview of new values derived in this paper from published
      results, compared to existing \CPT violation results. The new values for
      $\Dan-0.38\Daz$ in the \Bd and \Bs systems should be regarded as crude
      estimates as they are based on an estimate for the average \B momentum.}
  \label{tab:CPT_overview}
  \vspace{0.1cm}
  {\footnotesize
  \begin{tabular}{l l l@{} @{}r l@{} @{}r} \hline
System & Parameter & \multicolumn{2}{l}{Current best value} & 
\multicolumn{2}{l}{New value} \\ 
\hline
\Bd  & $\Re(z)$ & $(1.9\pm4.0)\%$ & \cite{Aubert:2004xga,Higuchi:2012kx} & 
       $(0.7\pm2.4)\%$ & \cite{HFAG}  \\
     & $N_B(\Dan-0.30\Daz)^\dagger$ & $(-3.0\pm2.4)\times10^{-15}\gev$ & \cite{Aubert:2007bp} &  &  \\
     & $\Dan-0.38\Daz$ &  &  & 
       $(0.9\pm2.8)\times10^{-15}\gev\;$ & \cite{LHCb-PAPER-2012-035}  \\
\hline
\Bs  & $\Re(z)$ & - & & $(6\pm4)\%$ & \cite{LHCb-PAPER-2013-002} \\
     & $\Dat^{\dagger\dagger}$ & $(3.7\pm3.8)\times10^{-12}\gev$ & 
       \cite{Kostelecky:2010bk} & &\\
     & $\Dan-0.38\Daz$ & & & $(5\pm3)\times10^{-14}\gev$ & 
       \cite{LHCb-PAPER-2013-002}  \\
\hline
\multicolumn{6}{l}{$^\dagger$ $N_B\equiv\DGd/\dmd$, which is about $1/190$ in the Standard Model.} \\
\multicolumn{6}{l}{$^{\dagger\dagger}$ \Dat is the constant component of \Damu that depends here on the orientation of \dzero.}
\end{tabular}
}
\end{center}
\end{table}

\begin{table}[!b]
  \begin{center}
    \caption{Expected statistical sensitivities (one standard deviation) on \CPT
      parameters with the existing $3\invfb$ data set from \lhcb using the
      listed decay modes, compared to current experimental limits. The
      uncertainties are expected to be dominated by the statistical uncertainty
      as in the current measurements.}
  \label{tab:CPT_reach}
  \vspace{0.1cm}
  {\small
  \begin{tabular}{l l l@{} @{}r l l } \hline
System & Parameter & \multicolumn{2}{l}{Current best limit} & \lhcb $3\invfb$ & Decay mode \\ 
\hline
\Dz  & $|\Re(z)y-\Im(z)x|$ & $(0.83\pm0.77)\%~^{\dagger}$ & \cite{Link:2002fg}~ & $0.02\%~^{\dagger}$ & $\Dz\to\Km\pip$ \\ 
     & \Damu         & $\sim3\times10^{-13}\gev$ & \cite{Link:2002fg}~ & $1\times 10^{-14}\gev$ & $\Dz\to\Km\pip$ \\ 
\hline
\Bd  & $\Im(z)$        & $(-0.8\pm0.4)\%$ & \cite{PDG2014}~ & $0.1\%$  & $\Bd\to\D^{(*)-}\mup\neum$ \\
     & $\Re(z)$       & $(1.9\pm4.0)\%$ & \cite{Aubert:2004xga,Higuchi:2012kx}~ & $7\%$    & $\Bd\to\jpsi\KS$ \\
     & \Damu       & $\order(10^{-13})\gev$ & \cite{Aubert:2007bp}~ & $1\times 10^{-15}\gev$ & $\Bd\to\jpsi\KS$ \\ 
\hline
\Bs  & $\Im(z)$       & -- & & $0.4\%$  & $\Bs\to\Dsm\pip$ \\
     & $\Re(z)$       & -- & & $2\%$    & $\Bs\to\jpsi\Pphi$ \\
     & \Damu        & $\order(10^{-12})\gev$ & \cite{Kostelecky:2010bk}~ & $1\times 10^{-14}\gev$ & $\Bs\to\jpsi\Pphi$\\
\hline
\multicolumn{6}{l}{$^{\dagger}$ Assuming that $x\approx y\approx0.5\%$.}\\
\end{tabular}
}
\end{center}
\end{table}
\renewcommand{\arraystretch}{1.0}

\subsection{Neutral kaons}

In the neutral kaon system, there are many experimental searches for \CPT
violation. Most of them have been performed within the classical framework,
i.e., assuming $z$ to be constant. In the PDG review~\cite{PDG2014}, combining
results from the KLOE, KTeV, CPLEAR and NA48 experiments, average values of
$\Re(\delta)=(2.4\pm2.3)\times10^{-4}$ and
$\Im(\delta)=(-0.7\pm1.4)\times10^{-5}$ are reported, where
$\delta\approx-z/2$. An experimental limit on direct \CPT violation is also
included in this review.

A search for sidereal variations in the SME framework has been performed at the
KLOE experiment~\cite{Babusci:2013gda}. The kaons are produced from the \Pphi
resonance, which has a small boost of $\beta\gamma\simeq0.015$, and detected in
the $\pip\pim$ final state. Limits on all four SME parameters are reported with
uncertainties on \Damu of about $2\times10^{-18}\gev$.\footnote{Natural units
  are used with $c=1$.} Another search for sidereal variations using KTeV data
is presented in Ref.~\cite{Nguyen:2001tg}, which has not been published in a
peer-reviewed journal. Due to the high boost of the uncorrelated kaons
($\beta\gamma\approx100$), strong limits on the sidereal-phase-dependent SME
parameters have been set to $\Daxy<9.2\times10^{-22}\gev$ at 90\% confidence
level (CL). Kaons produced at the E773 experiment are also highly boosted
($\beta\gamma\simeq100$). Using E773 results, a bound on the constant SME
parameters has been determined in Refs.~\cite{Kostelecky:1997mh,
  Kostelecky:2010bk} to $|\Dan-0.6\Daz|\lesssim5\times10^{-21}\gev$. Even though
cross sections for kaon and \Pphi production are high at the LHC, it will be
difficult for \lhcb to compete with the dedicated kaon experiments due to the
limited decay time acceptance (roughly up to one \KS lifetime), lower boost and
larger backgrounds.

\subsection{Neutral charm}

Only the FOCUS collaboration has reported limits on \CPT violation in \Dz
mixing~\cite{Link:2002fg}. About 35k Cabibbo-favoured $\Dz\to\Km\pip$
decays\footnote{The inclusion of charge-conjugated decay modes is implicit.}
have been analysed, both in the classical and SME approach. This final
state is not a pure flavour-specific eigenstate, since there is also a small
contribution from doubly Cabibbo-suppressed $\Dz\to\Kp\pim$ decays. Due to the
small mixing in the \Dz system~\cite{HFAG}, the \CPT asymmetry can be
approximated to first order as
\begin{align}
  \ACPT(t) = \Adir 
  - \sqrt{R_D}\sin\phi(x \cos\delta + y \sin\delta) \Gamma t
  + (\Re(z) y - \Im(z) x ) \Gamma t  \ , 
\label{eq:DzClassic}
\end{align}
where $x\equiv\Delta m/\Gamma$, $y\equiv\Delta\Gamma/2\Gamma$,
$R_D=(0.349\pm0.004)\%$~\cite{HFAG} is the decay rate ratio of doubly
Cabibbo-suppressed over Cabibbo-favoured decays and $\phi$ and $\delta$ are the
corresponding weak and strong phases. The second term, the contribution from \CP
violation, is maximally of $\order(10^{-4})$~\cite{HFAG} and is neglected in the
FOCUS analysis. In their classical analysis, a value of
$\Re(z)y-\Im(z)x=(0.83\pm0.77)\%$ is reported. Assuming $x\approx y\approx
0.5\%$, this measurement provides only a weak bound on $\Re(z) - \Im(z)$ of
$\order(1)$.

At \lhcb, many more $\Dz\to\Km\pip$ decays are available. In the current
$3\invfb$ data sample, more than 50M Cabibbo-favoured decays have been
observed~\cite{LHCb-PAPER-2013-053}, which means a possible improvement of the
FOCUS measurement by a factor of about $40$ and a precision on $\Re(z) y -
\Im(z) x$ of $0.02\%$. At this precision, the \CP-violating term cannot be
ignored anymore and needs to be taken into account in the analysis. In addition,
the observed \CPT asymmetry will include effects from production and detection
asymmetries. Fortunately, the latter two effects are expected to be independent
of the \Dz decay time, adding only to the constant contribution from direct \CP
violation, \Adir.

The same FOCUS paper~\cite{Link:2002fg} also presents a full sidereal analysis
in the SME framework. The average boost of the \Dz mesons is $\mean{\beta\gamma}
\approx 39$. Due to the SME constraint, the \CPT-violating term in
Eq.~\ref{eq:DzClassic} is zero and a further expansion in $x$ and $y$ is
required, which reduces the sensitivity to $\Re(z)$. The expansion to second and
third order in decay time gives
\begin{align}
   \ACPT(t) = \frac{\Re(z)(x^2 + y^2)(\Gamma t)^2}{2x} 
   \times\left[\frac{xy}{3}\Gamma t + 
     \sqrt{R_D}(x\cos\delta+y\sin\delta)\right] \ ,
\label{eq:DzSME}
\end{align}
where \Adir and all \CP-violating terms are omitted. Assuming again $x\approx
y\approx 0.5\%$, the uncertainties on the \Damu parameters are found to be about
$3\times10^{-13}\gev$. At \lhcb, with their large sample of $\Dz\to\Km\pip$
decays and assuming a comparable boost factor, it should be possible to improve
the FOCUS bounds by a factor 40. Note, however, that it will not be possible to
constrain $\Re(z)$ to be smaller than one, since $\Re(z)$ is suppressed in
Eq.~\ref{eq:DzSME} by $\order(10^{-6})$. Nevertheless, no assumptions on the
smallness of $|z|$ have been made so far. Extrapolating to the statistically
larger sample, \lhcb should be able to reach a sensitivity on the \Damu
parameters of about $1\times10^{-14}\gev$.

\subsection[B0 mesons]{\boldmath{\Bd} mesons}

Due to the small value of $\DGd$ in the \Bd system, decays to flavour-specific
final states are sensitive to $\Im(z)$, while decays to \CP eigenstates are
sensitive to $\Re(z)$ (cf. Eqs.~\ref{eq:ACPT_fs} and \ref{eq:ACPT_CP}). This is
a key point that is used below for \Bd decays, but it is also valid for \Bs
decays. Using only dilepton (i.e., flavour-specific) final states, the \babar
collaboration published $\Im(z)=(-1.39 \pm
0.80)\%$~\cite{Aubert:2006nf}. Similarly, the \belle collaboration reported
$\Im(z) = (-0.57\pm0.47)\%$, mainly using flavour-specific final
states~\cite{Higuchi:2012kx}. The average value of both results is
$\Im(z)=(-0.8\pm0.4)\%$~\cite{PDG2014}. Using the same dilepton final states,
the \babar collaboration also reported a measurement of $\Re(z)\DGd =
(-7.1\pm4.4) \times 10^{-3}\invps$~\cite{Aubert:2006nf}. When inserting the
theoretical expectation of 
$\DGd \approx -(2.7\pm0.7)\times10^{-3}\invps$~\cite{Lenz:2006hd},
this measurement gives only
a weak constraint on $\Re(z)$ of $\order(2)$. Since $|z|^2$ terms have been
ignored in this analysis, this means that a higher sensitivity to $\Re(z)$ could
have been achieved when including $|z|^2$ terms in the fits to the decay rates.

Due to the relatively low tagging performance in a hadron-collider environment,
the untagged asymmetry for flavour-specific decays, defined as
\begin{align}
   A_{\rm untagged}(t) &\equiv 
   \frac{\left[P_f,(t)+\Pbar_f(t)\right]-
     \left[P_{\fbar}(t)+\Pbar_{\fbar}(t)\right]}
        {\left[P_f,(t)+\Pbar_f(t)\right]+
          \left[P_{\fbar}(t)+\Pbar_{\fbar}(t)\right]} \ ,
\end{align}
gives a higher sensitivity to $\Im(z)$ than the tagged asymmetry as defined in
Eq.~\ref{eq:ACPT}. Including experimental effects from a possible detection
asymmetry \Ad and from a production asymmetry \Ap, the observed asymmetry
becomes
\begin{align}
   A_{\rm untagged}^{\rm observed}(t) &= \Ad + \Amix/2 - 
   \left(\Amix/2-\Ap\right)\cos\dmd t + \Im(z)\sin{\dmd t} \ ,
   \label{eq:ACPT_untagged}
\end{align}
whereby $|z|^2$ terms have been ignored and $\DGd$ is approximated to be
zero. Compared to Eq.~\ref{eq:ACPT_fs}, the sensitivity to $\Im(z)$ is only
reduced by a factor 2, rather than a reduction by a factor $20-30$, which is the
typical loss due to the flavour tagging in a hadronic environment. In
Eq.~\ref{eq:ACPT_untagged}, \Amix is the flavour-specific \CP asymmetry in \Bd
mixing. At \lhcb, using inclusive $\Bd\to\D^{(*)-}\mup\neum$ decays, a
high-precision measurement of $\Im(z)$ is possible, since the dilution of the
amplitude of the oscillation due to the partial reconstruction is
small~\cite{LHCb-PAPER-2013-036}. We estimate about 3 million inclusive
$\Bd\to\D^{(*)-}\mup\neum$ decays in the $3\invfb$ data set, using the observed
yields in $\Bs\to\Dsm\mup\neum$ decays~\cite{LHCb-PAPER-2013-033} and the
production ratio of \Bd and \Bs mesons~\cite{LHCb-PAPER-2012-037}. Hence, a
statistical precision on $\Im(z)$ of $0.1\%$ is in reach.

In \Bd decays to \CP final states, $\Re(z)$ appears in the cosine term of the
time-dependent \CP asymmetry. Neglecting \CP violation in mixing (i.e.,
$\Amix=0$), and setting $\Delta\Gamma_d=0$ and $\Im(z)=0$, the time-dependent
(tagged) asymmetry from Eq.~\ref{eq:ACPT_CP} becomes
\begin{align}
   \ACPTCP(t) = D_f\Re(z) - \left[ C_f + D_f \Re(z) \right] \cos\dmd t + S_f \sin\dmd t \ .
\label{eq:ACPT_CPfsBd}
\end{align}
Effects from $\Im(z)$ are expected to be negligible and this assumption can be
tested with experimental input from flavour-specific decay modes as described
above. Similarly, \Amix is also negligible at the current experimental
precision~\cite{HFAG}. Direct \CP violation ($C_f$) and $\Re(z)$ both contribute
to the cosine term. In principle, the time-independent offset is also sensitive
to $\Re(z)$, however, this offset is additionally affected by production,
detection and tagging asymmetries. Hence, in practice most information on
$\Re(z)$ will come from the oscillating term.

For $\Bd$ decays to the \CP final state $\jpsi\KS$, we can identify $C_f=0$,
$D_f=\cos2\beta$ and $S_f=\sin2\beta$, where $\beta$ is the usual CKM parameter.
We ignored for simplicity small effects coming from \CP violation in kaon and
\Bd mixing and direct \CP violation due to the penguin contributions. The
contribution from direct \CP violation gives the dominant uncertainty on $C_f$
and therefore on the determination of $\Re(z)$. Theoretically, it is estimated
to be at most a few times $10^{-3}$~\cite{Grossman:2002bu}. Experimentally, the
direct \CP asymmetry in $\Bp\to\jpsi\Kp$ decays is
$(0.3\pm0.6)\%$~\cite{PDG2014}, which is expected to be largely equal to that in
$\Bd\to\jpsi\KS$ decays using isospin symmetry~\cite{Fleischer:2001cw}. Another
experimental constraint comes from the $\Bd\to\jpsi\piz$ decay, which can be
used to determine the direct \CP violation in $\Bd\to\jpsi\KS$ to be
$(1\pm1)\%$~\cite{Faller:2008zc}.

The \belle collaboration has measured $\Re(z) = (1.9\pm5.0)\%$, where the
sensitivity mainly comes from $\Bd\to\jpsi\KSL$
decays~\cite{Higuchi:2012kx}. Similarly, the \babar collaboration has measured
with a small fraction of the data $\Re(z)\Re(\lambda)/|\lambda| = (1.4 \pm
4.9)\%$~\cite{Aubert:2004xga}. We can remove the factor
$\Re(\lambda)/|\lambda|=D_f=\cos(2\beta)=0.722^{+0.016}_{-0.020}$, where we
used the measured value of the CKM angle $\beta$ from
Ref.~\cite{CKMfitter}. Then, this measurement translates to $\Re(z) =
(1.9\pm6.8)\%$, where the uncertainty from the factor $\Re(\lambda)/|\lambda|$
is negligible. This result was left unnoticed in the PDG world average of
$\Re(z)$~\cite{PDG2014}. Averaging here both numbers, we find
$\Re(z)=(1.9\pm4.0)\%$. Both results neglect the possible contribution from
direct \CP violation. A more recent and accurate value on $\Re(z)$ can actually
be obtained using the world average on the cosine coefficient of $(0.5 \pm
1.7)\%$~\cite{HFAG}. With $\Re(\lambda)/|\lambda| \approx 0.72$ and setting
$C_f=0$, this results in $\Re(z)=(0.7\pm2.4)\%$.

Finally, we briefly mention the \babar analysis~\cite{Lees:2012uka} where the
\CP, \CPT and \T asymmetries are tested separately. For instance, \CPT
asymmetries for \Bd mixing are constructed by simultaneously interchanging the
time ordering of initial \Bz and \Bzb decays and substituting \KL and \KS
states. Although this method is statistically not competitive, it does allow to
cleanly separate effects from \CP, \T and \CPT violation. Unfortunately, such
tests are only possible at the \FourS\ experiments, and not at hadron collider
experiments where the \Bd mesons are produced incoherently and the
reconstruction of \KL mesons is much more challenging.

In the SME framework, due to the constraint $\Re(z)\DGd \approx 2\Im(z)\dmd$,
the real part of $z$ is about 380 times larger than the imaginary part. Without
loss of generality, we used here the theoretical expectation value of $\DGd$
from Ref.~\cite{Lenz:2006hd}; experimentally $2\dmd/\DGd$ is already
bounded to be larger than 77 at 95\% CL~\cite{PDG2014}.  Therefore, \Bd decays
to \CP eigenmodes are more sensitive to the \Damu variables than \Bd decays to
flavour-specific modes. The \babar collaboration published a
paper~\cite{Aubert:2007bp} where the \Damu parameters are determined in a full
sidereal analysis. The boost of the \Bd mesons is $\beta\gamma=0.55$. They used
only dilepton events, rather than \CP eigenmodes. Using the expected $\Delta
\Gamma_d$ value~\cite{Lenz:2006hd}, the uncertainties on the \Damu parameters
are $\sim(5-25)\times10^{-13}\gev$, corresponding to uncertainties on $\Re(z)$
of order one. Just as in their classical analysis~\cite{Aubert:2006nf}, this
means that a higher sensitivity to \Damu would have been possible in case
$|z|^2$ terms are not ignored.

When using the location and orientation of the \babar and \belle experiments and
their measurements of $\Re(z)$, stronger constraints on the constant \Damu term
can be set, however, at this point we focus on \lhcb where an even higher
precision can be reached due to the larger boost of the \Bd mesons. The average
momentum of $b$ hadrons at \lhcb is $\mean{p} \approx
80\gev$~\cite{Altarelli:2008xy}, corresponding to a relativistic boost of
$\mean{\beta\gamma}\approx15$. The \lhcb collaboration reported a value of
$C_f=(3\pm9)\%$ using $\Bd\to\jpsi\KS$ decays in the $1\invfb$ data
set~\cite{LHCb-PAPER-2012-035}. This corresponds to $\Re(z)=(4\pm12)\%$.  Using
the \lhcb beam direction, a measurement of the constant combination of SME
parameters of $(\Dan-0.38\Daz) = (0.9\pm2.8)\times10^{-15}\gev$ is
obtained. Although this number is only a crude estimate, mainly due to the
uncertainty on the average \Bd momentum, it improves the current best
value~\cite{Aubert:2007bp} by two orders of magnitude.  By making use of the \Bd
momentum in each event and with a full sidereal analysis on the $3\invfb$ data
set, \lhcb should be able to reach a sensitivity on \Damu of about
$1\times10^{-15}\gev$.

\subsection[Bs mesons]{\boldmath{\Bs} mesons}

The discussion for the \Bs system is very similar to that for the \Bd system.
In this system $\DGs$ is not anymore negligible, but still small enough such
that flavour-specific final states primarily give access to $\Im(z)$, while \CP
eigenmodes give access to $\Re(z)$. No dedicated \CPT measurements have been
done with \Bs mesons to date. In the classical approach, \lhcb would be able to
measure $\Im(z)$ using the flavour-specific $\Bs\to\Dsm\pip$ decays. In the
$3\invfb$ data set, $N=100$k untagged signal decays can be
expected~\cite{LHCB-PAPER-2013-006}. Following Eq.~\ref{eq:ACPT_untagged}, this
corresponds to a statistical uncertainty on $\Im(z)$ of $\sqrt{2/N}=0.4\%$ (see
Table~\ref{tab:CPT_reach}). Alternatively, also the more abundant inclusive
$\Bs\to\Dsm\mup\neum$ decays can be used to measure $\Im(z)$. Due to the partial
reconstruction, however, the worse time resolution washes out the oscillations
already after a $1\ps$ (see Ref.~\cite{LHCb-PAPER-2013-036}), reducing the
sensitivity.

Constraints on $\Re(z)$ can be made using \Bs decays to the \CP eigenstate
$\jpsi\Pphi$. This decay mode is the \Bs equivalent of $\Bd\to\jpsi\KS$.
Equation~\ref{eq:ACPT_CP} gives the observable asymmetry. The phase
$\arg(\lambda_f)=\phi_s$ is expected~\cite{CKMfitter} and experimentally
measured~\cite{HFAG} to be small, leading to $D_f\approx1$ and
$S_f\approx0$. Any effect from \Amix can be ignored at the current level of
precision~\cite{HFAG}. The \lhcb collaboration has published a value of
$|\lambda_f|=0.94\pm0.04$ using the $1\invfb$ data
set~\cite{LHCb-PAPER-2013-002}. Ignoring again direct \CP violation, a first
evaluation of $\Re(z)\approx(1-|\lambda_f|^2)/2=(6\pm4)\%$ can be made in the
\Bs system.

In the \Bs system, the SME constraint $\Re(z)\DGs \approx 2\dms \Im(z)$ leads to
a $\Re(z)$ that is a factor 450 larger than $\Im(z)$. Even more than for the \Bd
system, this means that one should focus on decays to \CP final states, such as
$\Bs\to\jpsi\Pphi$. An interesting relation between the \Kz, \Bd and \Bs systems
is pointed out in Ref.~\cite{Kostelecky:2010bk}. As the expectation value of
\Damu is dominated by the valence quarks, a sum rule relating these three
neutral meson systems can be written as
\begin{align}
  \DamuK - \DamuBd + \DamuBs \approx 0 \ .
\end{align}
Since the constraints on \DamuK are most strong and compatible with zero, this
sum rule implies that possible \CPT-violating effects in the \Bd and \Bs system
should be of the same order. In that sense, the \Bd system is more interesting,
since the production rate of \Bd mesons is higher and the mass difference \dmd
is smaller (c.f. Eq.~\ref{eq:SMEz}). Ideally, the mass difference should be such
that one could just measure one period of oscillation, which is the case for \Bd
oscillations.  Nevertheless, it remains important to measure possible \CPT
violation in all possible systems to verify this sum rule.

Using the like-sign dimuon asymmetry measured in the \dzero data, a value for
\Damu has been derived in Ref.~\cite{Kostelecky:2010bk}. Assuming that the only
source of \CPT violation comes from \Bs decays (like-sign dimuons originate from
both \Bd and \Bs mixing) and using the average boost of
$\mean{\beta\gamma}=4.1$, the constant \Damu term becomes
$(3.7\pm3.8)\times10^{-12}\gev$. This corresponds to
$\Re(z)=1.0\pm0.8$. Stronger limits on $\Re(z)$ can be set with the \CP
eigenmode decay $\Bs\to\jpsi\Pphi$. Using again $\Re(z)=(6\pm4)\%$, derived from
\lhcb results~\cite{LHCb-PAPER-2013-002}, and taking as average boost
$\mean{\beta\gamma}\approx15$, we find as a crude estimate
$(\Dan-0.38\Daz)=(5\pm3)\times10^{-14}\gev$, which is an improvement by two
orders of magnitude. With the existing \lhcb data set and a dedicated sidereal
analysis, it should be possible to reach a sensitivity of about
$1\times10^{-14}\gev$ or below.

\section{Conclusion}
\label{sec:conclusion}

We have presented new results on \CPT violation in \Bd and \Bs mixing in both
the classical and SME approach, derived from published \babar, \belle and \lhcb
results. The new results in the SME approach should be regarded a crude
estimates, as a precise estimate of the average \B momentum is missing. In both
approaches there is a significant improvement over previous results (see
Table~\ref{tab:CPT_overview}). \lhcb should be able to further improve these
numbers in the \Bd and \Bs systems, as well as in the \Dz system, with dedicated
analyses on the existing $3\invfb$ data set (see Table~\ref{tab:CPT_reach}). In
most cases these possible \lhcb measurements would improve the current best
values by orders of magnitude and the corresponding precision on \Damu is
approaching the interesting region of $m^2/M_{\rm Pl}$. Further improvements can
be expected with the \lhcb data from run II, starting in 2015. On the longer
time scale, much stronger limits can be expected from \belle II and the \lhcb
upgrade.

\section*{Acknowledgements}

\noindent We thank Hans Wilschut, Gerco Onderwater, Keri Vos and other
colleagues from the University of Groningen for the fruitful discussions on
Lorentz invariance violation. We thank Marcel Merk for his feedback on the paper
draft, Robert Fleischer for his comments on the possible contribution from
direct \CP violation, and Patrick Koppenburg and Guy Wilkinson for useful
suggestions on the text. This work is supported by the Netherlands Organisation
for Scientific Research (NWO Vidi grant 680-47-523) and the Foundation for
Fundamental Research on Matter (FOM).

\addcontentsline{toc}{section}{References}
\setboolean{inbibliography}{true}
\bibliographystyle{LHCb}
\bibliography{main}

\end{document}